\documentclass[aps,prl,twocolumn,showpacs,superscriptaddress]{revtex4}
\usepackage{amsmath}
\usepackage{amssymb}
\usepackage{epsfig}
\usepackage{color}
\usepackage{bm}

\begin{document}
\title{Nature of the quantum critical point as
disclosed by extraordinary behavior of magnetotransport and the
Lorentz number in the heavy-fermion metal $\rm \bf YbRh_2Si_2$}
\author{V. R. Shaginyan}\email{vrshag@thd.pnpi.spb.ru}
\affiliation{Petersburg Nuclear Physics Institute, Gatchina,
188300, Russia}\affiliation{Clark Atlanta University, Atlanta, GA
30314, USA} \author{A. Z. Msezane}\affiliation{Clark Atlanta
University, Atlanta, GA 30314, USA}\author{K. G. Popov}\affiliation{Komi
Science Center, Ural Division, RAS,
Syktyvkar, 167982, Russia}
\author{J.~W.~Clark}
\affiliation{McDonnell Center for the Space Sciences \& Department
of Physics, Washington University, St.~Louis, MO 63130, USA}
\author{M.~V.~Zverev}
\affiliation{Russian Research Centre Kurchatov Institute, Moscow,
123182, Russia} \affiliation{Moscow Institute of Physics and
Technology, Moscow, 123098, Russia}
\author{V. A. Khodel} \affiliation{Russian Research Centre Kurchatov Institute,
Moscow, 123182, Russia} \affiliation{McDonnell Center for the Space
Sciences \& Department of Physics, Washington University, St.
Louis, MO 63130, USA}

\begin{abstract}  Physicists are engaged in vigorous debate on
the nature of the quantum critical points (QCP) governing the
low-temperature properties of heavy-fermion (HF) metals.
Recent experimental observations of the much-studied compound
$\rm YbRh_2Si_2$ in the regime of vanishing temperature
incisively probe the nature of its magnetic-field-tuned
QCP.  The jumps revealed both in the residual resistivity $\rho_0$
and the Hall resistivity $R_H$, along with violation of the
Wiedemann-Franz law, provide vital clues to the origin of such
non-Fermi-liquid behavior. The empirical facts point unambiguously
to association of the observed QCP with a fermion-condensation phase
transition.  Based on this insight, the resistivities $\rho_0$ and $R_H$
are predicted to show jumps at the crossing of the QCP produced
by application of a magnetic field, with attendant violation of the
Wiedemann-Franz law.  It is further demonstrated that experimentally
identifiable multiple energy scales are related to the scaling
behavior of the effective mass of the quasiparticles responsible
for the low-temperature properties of such HF metals.
\end{abstract}

 \maketitle

A quantum critical point (QCP) dictates the non-Fermi liquid (NFL)
low-temperature properties of strongly correlated Fermi systems,
notably heavy fermion (HF) metals, high-temperature
superconductors, and quasi-two-dimensional $^3$He. Their NFL
behavior is so radical that the traditional Landau quasiparticle
paradigm is at a loss to describe it. The underlying nature of the
QCP has continued to defy theoretical understanding.  Attempts have
been made using concepts such as the Kondo lattice and involving
quantum and thermal fluctuations at the
QCP\cite{stew,loh,si,sach,col}. Alas, when designed to describe one
property deemed central, these approaches fail to explain others,
even the simplest ones such as the Kadowaki-Woods relation
\cite{kadw,shagrep}. This relation, which emerges naturally when
quasiparticles of effective mass $M^*$ play the main role, can
hardly be explained within the framework of a theory that
presupposes the absence of quasiparticles at the QCP (for recent
reviews see \cite{shagrep,shag,mig100}).  Arguments that
quasiparticles in strongly correlated Fermi liquids ``get heavy and
die'' at the QCP commonly employ the assumption that the
quasiparticle weight factor $Z$ vanishes at the point of an
associated second-order phase transition \cite{col1,col2}. However,
this scenario is problematic \cite{khodz,clark10}.  Numerous
experimental facts have been discussed in terms of such a
framework, but how it can explain the physics of HF metals
quantitatively is left as an open question \cite{shagrep}. A theory
of fermion condensation (FC) that preserves quasiparticles while
being intimately related to the unlimited growth of $M^*$ has been
proposed and developed \cite{khs,volovik,shagrep,shag,mig100}.
Extensive studies have shown that this theory delivers an adequate
theoretical explanation of the great majority of experimental
results in different HF metals. In contrast to the Landau paradigm
based on the assumption that $M^*$ is a constant, within FC theory
$M^*$ depends strongly on both temperature $T$ and imposed magnetic
field $B$.  Accordingly, an extended quasiparticle paradigm is
introduced. The essential point is that -- as before --
well-defined quasiparticles determine the thermodynamic and
transport properties of strongly correlated Fermi systems, while
the dependence of the effective mass $M^*$ on $T$ and $B$ gives
rise to the observed NFL behavior \cite{shagrep,shag,mig100}.  The
most fruitful strategy for exploring and revealing the nature of
the QCP is to focus on those properties that exhibit the most
spectacular departures from Landau Fermi Liquid (LFL) behavior in
the zero-temperature limit.  In particular, incisive experimental
measurements recently performed on the heavy-fermion metal $\rm
YbRh_2Si_2$ have probed the nature of its magnetic-field-tuned QCP.
It is found that at vanishingly low temperatures the residual
resistivity $\rho_0$ experiences a jump across the magnetic QCP,
with a crossover regime proportional to $T$
\cite{steg1,steg2,steg3,steg_cm}. Jumps of the magnetoresistivity,
the Hall coefficient, and the Lorenz number at zero temperature are
in conflict with the common behavior of Kondo systems, for which
the width of the change remains finite at zero temperature
\cite{steg3,s_k}.  Under the same experimental conditions in $\rm
YbRh_2Si_2$, the Hall coefficient $R_H$ is also found to experience
a jump \cite{steg2}, while the data collected on heat and charge
transport at the QCP can be interpreted as indicating a violation
of the Wiedemann-Franz law \cite{steg3}.  The Wiedemann-Franz law
defines the value of the Lorentz number $L=\kappa/T\sigma$ at
$T\to0$, i.e., $L=L_0$ with $L_0=(\pi k_B)^2/3e^2$, where $\kappa$,
$\sigma$, $k_B$, and $e$ are respectively the thermal conductivity,
the electrical conductivity, Boltzmann's constant, and the charge
of the electron.

In this Letter we study magnetotransport and violation of the
Wiedemann-Franz law in $\rm YbRh_2Si_2$ across a QCP tuned by
application of a magnetic field. Close similarity between the
behavior of the Hall coefficient $R_H$ and magnetoresistivity
$\rho$ at QCP indicates that all manifestations of magnetotransport
stem from the same underlying physics.  We show that the violation
of the Wiedemann-Franz law together with the jumps of the Hall
coefficient and magnetoresistivity in the zero-temperature limit
provide unambiguous evidence for interpreting the QCP in terms of a
fermion condensation quantum phase transition (FCQPT) forming a
flat band in $\rm YbRh_2Si_2$.

\begin{figure}[!ht]
\begin{center}
\includegraphics [width=0.48\textwidth]{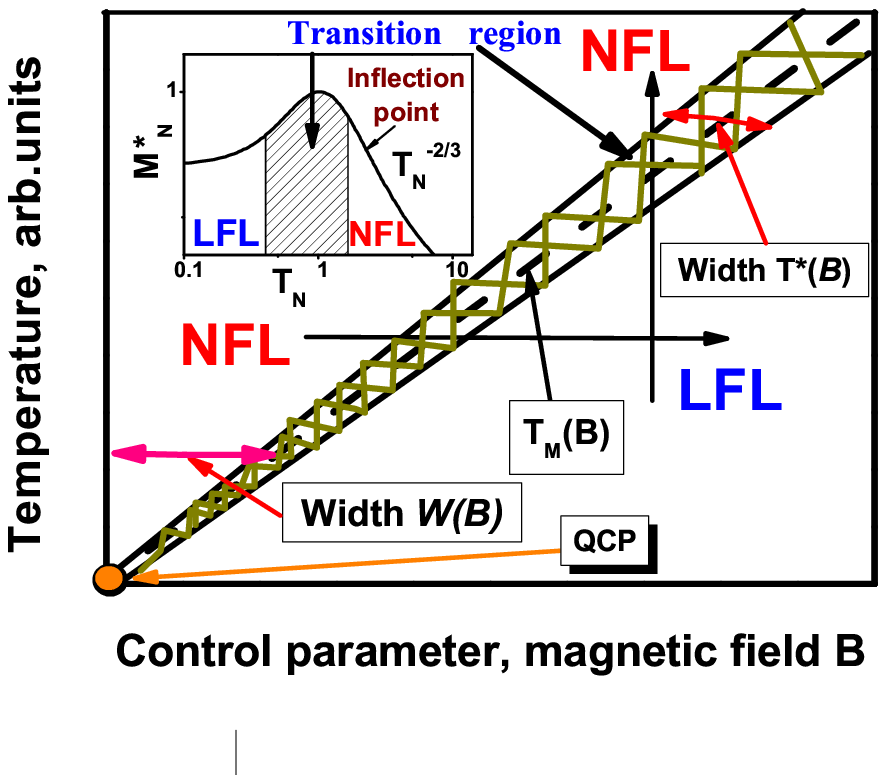}
\vspace*{-1.5cm}
\end{center}
\caption{(color online). Schematic $T-B$ phase diagram of HF liquid
with magnetic field as the control parameter. The vertical and
horizontal arrows show LFL-NFL and NFL-LFL transitions at fixed $B$
and $T$, respectively. At $B=0$ the system is in its NFL state
having a flat band and demonstrates NFL behavior down to $T\to0$.
The hatched area separates the NFL phase and the weakly polarized LFL
phase and represents the transition area. The dashed line in the hatched
area represents the function $T_M(B)$ given by Eq.~\eqref{TMB}. The
functions $W(B)\propto T$ and $T^*(B)\propto T$ shown by two-headed
arrows define the width of the NFL state and the transition
area, respectively. The QCP located at the origin and
indicated by an arrow denotes the critical point at which the
effective mass $M^*$ diverges and both $W(B)$ and $T^*(B)$ tend to
zero. The inset shows a schematic plot of the normalized effective mass
versus the normalized temperature.  The transition regime, where $M^*_N$
reaches its maximum value at $T_N=T/T_M=1$, is shown as the hatched
area in both the main panel and the inset.  Arrows indicate
the transition region and the inflection point $T_{\rm inf}$ in
the $M^*_N$ plot.}\label{fig1}
\end{figure}

We begin with an analysis of the scaling behavior of the effective
mass $M^*$ and $T-B$ phase diagram of a homogeneous HF liquid,
thereby avoiding complications associated with the crystalline
anisotropy of solids \cite{shagrep}.  Near the FCQPT, the
temperature and magnetic field dependence of the effective mass
$M^*(T,B)$ is governed by the Landau equation \cite{trio}
\begin{eqnarray}
\nonumber \frac{1}{M^*_{\sigma}(T,
B)}&=&\frac{1}{m}+\sum_{\sigma_1}\int\frac{{\bf p}_F{\bf
p}}{p_F^3}F_
{\sigma,\sigma_1}({\bf p_F},{\bf p}) \\
&\times&\frac{\partial n_{\sigma_1} ({\bf
p},T,B)}{\partial{p}}\frac{d{\bf p}}{(2\pi)^3}. \label{HC1}
\end{eqnarray}
where $F_{\sigma,\sigma_1}({\bf p_F},{\bf p})$ is the Landau
interaction, $p_F$ is the Fermi momentum, and $\sigma$ is the spin label.
To simplify matters, we ignore the spin dependence of the effective
mass, noting that $M^*(T,B)$ is nearly independent of spin in weak
fields. The quasiparticle distribution function $n$ can be
expressed as
\begin{equation} n_{\sigma}({\bf p},T)=\left\{ 1+\exp
\left[\frac{(\varepsilon({\bf p},T)-\mu_{\sigma})}T\right]\right\}
^{-1},\label{HC2}
\end{equation}
where $\varepsilon({\bf p},T)$ is the single-particle (sp)
spectrum. In the case being considered, the sp spectrum depends on
spin only weakly. However, the chemical potential $\mu_{\sigma}$
depends non-trivially on spin due to the Zeeman splitting,
$\mu_{\pm}=\mu\pm B\mu_B$, where $\pm$ corresponds to states with
the spin ``up'' or ``down.''  Numerical and analytical solutions of
this equation show that the dependence $M^*(T,B)$ of the effective
mass gives rise to three different regimes with increasing
temperature. In the theory of fermion condensation, if the system
is located near the FCQPT on its ordered side, then the fermion
condensate (FC) represents a group of sp states with dispersion
given by  \cite{noz}
\begin{equation}
\varepsilon({\bf p},n)-\mu=T\ln {1-n({\bf p})\over n({\bf p})},
\label{tem}
\end{equation}
where $\mu$ is the chemical potential and $n({\bf p})$ is the
quasiparticle occupation number, which loses its temperature
dependence at sufficiently low $T$. On the ordered side the sp
spectrum of the HF liquid contains a flat portion embracing the
Fermi surface; on the other hand, on the disordered side, at fixed,
finite $B$ and low temperatures we have a LFL regime with
$M^*(T)\simeq M^*+aT^2$, where $a$ is a positive constant
\cite{shagrep}. Thus the effective mass grows as a function of $T$,
reaching its maximum $M^*_M$ at some temperature $T_M$ and
subsequently diminishing according to \cite{ckhz}
\begin{equation}\label{r2}
M^*(T) \propto T^{-2/3}.
\end{equation}
Moreover, the closer the control parameter $B$ is to its critical
value $B_{c0}=0$, the higher the growth rate. In this case, the peak
value of $M^*_M$ also grows, but the temperature $T_M$ at which
$M^*$ reaches its peak value decreases, so that $M^*_M(T_M,B\to
B_{c0})\to\infty$. At $T>T_M$, the last traces of LFL disappear.
When the system is in the vicinity of the FCQPT, the approximate
interpolative solution of Eq.~\eqref{HC1} reads \cite{shagrep}
\begin{equation}
\frac{M^*}{M^*_M}={M^*_N(T_N)}\approx
c_0\frac{1+c_1T_N^2}{1+c_2T_N^{8/3}}. \label{fin1}
\end{equation}
Here, $T_N=T/T_M$ is the normalized temperature, with
$c_0=(1+c_2)/(1+c_1)$ in terms of fitting parameters
$c_1$ and $c_2$.  Since the magnetic field enters Eq.~\eqref{HC2} in
the form $\mu_BB/T$, we conclude that
\begin{equation}\label{YTB}
T/T_M\propto \frac{T}{\mu_BB},
\end{equation}
where $\mu_B$ is the Bohr magneton. It follows from Eq.~\eqref{YTB}
that
\begin{equation}
\label{TMB}
T_M\simeq a_1\mu_BB.
\end{equation}
Equation~\eqref{fin1} reveals the scaling behavior of the
normalized effective mass $M^*_N(T_N)$: values of the effective
mass $M^*(T,B)$ at different magnetic fields $B$ merge into a
single mass value $M^*_N$ in terms of the normalized variable
$T_N=T/T_M$ \cite{shagrep}.  The inset in Fig.~\ref{fig1}
demonstrates the scaling behavior of the normalized effective mass
$M^*_N$ versus the normalized temperature $T_N$.  The LFL phase
prevails at $T\ll T_M$, followed by the $T^{-2/3}$ regime at $T
\gtrsim T_M$.  The latter phase is designated as NFL due to the
strong dependence of the effective mass on temperature.  The
temperature region $T\simeq T_M$ encompasses the transition between
the LFL regime with almost constant effective mass and the NFL
behavior described by Eq.~\eqref{r2}.  Thus $T\sim T_M$ identifies
the transition region featuring a crossover between LFL and NFL
regimes. The inflection point $T_{\rm inf}$ of $M_N^*$ versus $T_N$
is depicted by an arrow in the inset of Fig.~\ref{fig1}.

The transition (crossover) temperature $T_M(B)$ is not actually the
temperature of a phase transition.  Its specification is necessarily
ambiguous, depending as it does on the criteria invoked for determination
of the crossover point.  As usual, the temperature $T^*(B)$ is extracted
from the field dependence of charge transport, for example from the
resistivity $\rho(T)$ given by
\begin{equation}
\rho(T)=\rho_0+AT^{\alpha_R},\label{res}
\end{equation}
where $\rho_0$ is the residual resistivity and $A$ is a
$T$-independent coefficient. The term $\rho_0$ is ordinarily
attributed to impurity scattering. The LFL state is characterized
by the $T^{\alpha_R}$ dependence of the resistivity with index
$\alpha_R=2$. The crossover (through the transition regime shown as
the hatched area in both Fig.~\ref{fig1} and its inset) takes
place at temperatures where the resistance starts to deviate from
LFL behavior, with the exponent $\alpha_R$ shifting from 2
into the range  $1<\alpha_R<2$.
The schematic phase diagram of a HF metal is depicted in Fig.~\ref{fig1},
with the magnetic field $B$ serving as the control parameter.  At $B=0$,
the HF liquid acquires a flat band corresponding to a strongly degenerate
state.  The NFL regime reigns at elevated temperatures and fixed magnetic
field.  With increasing $B$, the system is driven from the NFL region to
the LFL domain.  As shown in Fig.~\ref{fig1}, the system moves from
the NFL regime to the LFL regime along a horizontal arrow, and from
the LFL to NFL along a vertical arrow.  The magnetic-field-tuned QCP
is indicated by an arrow and located at the origin of the phase diagram,
since application of a magnetic field destroys the flat band and
shifts the system into the LFL state \cite{shagrep,shag,mig100}.
The hatched area denoting the transition region separates the NFL state
from the weakly polarized LFL state and contains the dashed line
tracing $T_M(B)$.  Referring to Eq.~\eqref{TMB}, this line is defined
by the function $T=a_1\mu_BB$, and the width $W(B)$ of the NFL state
is seen to be proportional $T$. In the same way, it can be shown that the
width $T^*(B)$ of the transition region is also proportional to
$T$.

\begin{figure}[!ht]
\begin{center}
\includegraphics [width=0.47\textwidth]{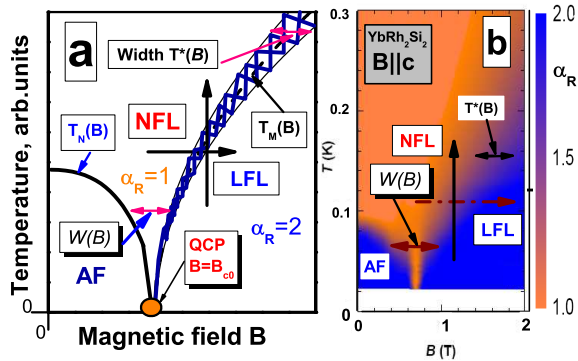}
\end{center}
\caption{(color online). Panel {\bf a} represents a schematic phase
diagram of the HF metal $\rm YbRh_2Si_2$, with $T_N(B)$ denoting
the N\`eel temperature as a function of magnetic field $B$.  The
QCP, identified by an arrow, is now shifted to $B=B_{c0}$.  At
$B<B_{c0}$ the system is in its antiferromagnetic (AF) state,
denoted by $\rm {AF}$. As in Fig.~\ref{fig1}, the vertical and
horizontal arrows show the transitions between the LFL and NFL
states, the functions $W(B)\propto T$ and $T^*(B)\propto T$
indicated with bi-directional arrows define the width of the NFL
state and of the transition region, respectively, and the dashed
line in the hatched area represents the function $T_M(B)$ given by
Eq.~\eqref{TMB}. The exponent $\alpha_R$ determines the
temperature-dependent part of the resistivity (cf.\
Eq.~\eqref{res}), with $\alpha_R$ taking values 2 and 1,
respectively, in LFL and NFL states.  In the transition regime the
exponent evolves between LFL and NFL values.  Panel {\bf b} shows
the experimental $T-B$ phase diagram \cite{cust,steg3}. The
evolution of $\alpha_R$ is depicted by color (coded in the vertical
stripe on the right-hand side of the panel). The NFL behavior
reaches to the lowest temperatures right at the QCP tuned by the
magnetic field.  The transition regime between the NFL state and
the field-induced LFL state broadens with rising magnetic fields
$B>B_{c0}$ and $T\sim T^*(B)$.  As in panel {\bf a}, transitions
from LFL to NFL state and from NFL to LFL state are indicated by
the corresponding arrows, as are $W(B)\propto T$ and $T^*(B)\propto
T$.} \label{fig2}
\end{figure}

In this letter we focus on the HF metal $\rm YbRh_2Si_2$, whose
empirical $T-B$ phase diagram is reproduced in panels {\bf a} and
{\bf b} of Fig.~\ref{fig2}.  Panel {\bf a} is similar to the main
panel of Fig.~\ref{fig1}, but with the distinction that this HF
compound possesses a finite critical magnetic field $B_{c0}\neq 0$
that shifts the QCP from the origin. To avert realization of a
strongly degenerate ground state induced by the flat band, the FC
must be completely eliminated at $T\to 0$. In a natural scenario,
this occurs by means of an antiferromagnetic (AF) phase transition
with an ordering temperature $T_N=70$ mK, while application of a
magnetic field $B=B_{c0}$ destroys the AF state at $T=0$
\cite{geg}. In other words, the field $B_{c0}$ places the HF metal
at the magnetic-field-tuned QCP and nullifies the N\`eel
temperature $T_{N}(B_{c0})=0$ of the corresponding AF phase
transition \cite{shagrep,shag4}. Imposition of a magnetic field
$B>B_{c0}$ drives the system to the LFL state. Thus, in the case of
$\rm YbRh_2Si_2$, the QCP is shifted from the origin to $B=B_{c0}$.
In FC theory, the quantity $B_{c0}$ is a parameter determined by
the properties of the specific heavy-fermion metal. In some cases,
notably the HF metal $\rm CeRu_2Si_2$, $B_{c0}$ does
vanish\cite{takah}, whereas in $\rm YbRh_2Si_2$, $B_{c0}\simeq
0.06$ T, $B\bot c$ \cite{geg}.

Panel {\bf b} of Fig.~\ref{fig2} portrays the experimental $T-B$
phase diagram in a manner showing the evolution of the exponent
$\alpha_R(T,B)$ \cite{cust,steg3}.  At the critical field
$B_{c0}\simeq 0.66$ T ($B\| c$), the NFL behavior extends down to
the lowest temperatures, while $\rm YbRh_2Si_2$ transits from the
NFL to LFL behavior under increase of the applied magnetic field.
Vertical and the horizontal arrows indicate, respectively, the
transition from the LFL to the NFL state and its reversal. The
functions $W(B)\propto T$ and $T^*(B)\propto T$ associated with
bi-directed arrows define the width of the NFL state and
transition region, respectively.  It noteworthy that the schematic
phase diagram displayed in panel {\bf a} of Fig.~\ref{fig2} is
in close qualitative agreement with its experimental counterpart
in panel {\bf b}.

To calculate the low-temperature dependence of $\rho$ on the
imposed magnetic field $B$ in the normal state of $\rm YbRh_2Si_2$,
we employ a model of a HF liquid possessing a flat band with
dispersion given by Eq.~\eqref{tem}.  Since the resistivity at
$T\to0$ is our primary concern, we concentrate on a special
contribution to the residual resistivity $\rho_0$ which we call the
critical residual resistivity $\rho_{0}^c$.  Analysis begins with
the case $B=0$, for which the resistivity of the HF liquid at low
temperatures is a linear function of $T$ \cite{shagrep,khodr}. This
observation is in accord with experimental facts derived from
measurements on $\rm YbRh_2Si_2$ indicating the presence of a flat
band in $\rm YbRh_2Si_2$ \cite{geg,shagrep,khodr,stegc}. In that
case, the effective mass $M^*(T)$ of the FC quasiparticles takes
the form
\begin{equation}
M^*(T)\sim
\frac{\eta p_F^2}{4T},
\label{M*}
\end{equation}
where $\eta=\delta p/p_F$ is determined by the characteristic size
$\delta p$ of the momentum interval $L$ occupied by the FC
\cite{arch}. With the result \eqref{M*} the width $\gamma$ of FC
quasiparticles is calculated in closed form, $\gamma\sim
\gamma_0+\eta T$, where $\gamma_0$ is a constant \cite{arch}. This
result leads to the lifetime $\tau_q$ of quasiparticles
\begin{equation}\label{LT}
\frac{\hbar}{\tau_q}\simeq a_1+a_2T,
\end{equation}
where $\hbar$ is Planck's constant, $a_1$ and $a_2$ are parameters.
Equation \eqref{LT} is in excellent agreement with experimental
observations \cite{tomph}. In general the electronic liquid in HF
metals is represented by several bands occupied by quasiparticles
that simultaneously intersect the Fermi surface, and FC
quasiparticles never cover the entire Fermi surface.  Thus there
exist LFL quasiparticles with the effective mass $M^*_L$
independent of $T$ and FC quasiparticles with $M^*$ given by
Eq.~\eqref{M*} at the Fermi surface, and all of them possess the
same width $\gamma$.  Upon appealing to the standard equation
\begin{equation}
\sigma\sim \frac{Ne^2}{\gamma M^*}
\label{sigma}
\end{equation}
for the conductivity $\sigma$ (see e.g. \cite{trio})
and taking into account the formulas specifying $M^*$ and $\gamma$,
we find that  $\sigma\sim Ne^2/(p_F\eta)^2$, where $N$ is the
number density of electrons. With this result, we arrive at a
critical residual resistivity $\rho_{0}^c$ that is independent of
$T$:
\begin{equation}
\rho_{0}^c\sim \frac{\eta^2}{p_Fe^2}. \label{rho}
\end{equation}
Careful derivation and examination of Eqs.~\eqref{sigma} and
\eqref{rho} is provided in \cite{arch}.  The term ``residual
resistivity'' is ordinarily attributed to impurity scattering. In
the present case, Eq.~\eqref{rho} shows that $\rho_0^c$ is
determined by the presence of a flat band and has no relation to
the scattering quasiparticles by impurities.

We next demonstrate that the application of a magnetic field to the
HF liquid generates the observed step-like drop in the residual resistivity
$\rho_0$.  Indeed, Fig.~\ref{fig1} informs us that at fixed temperature,
application of the field $B$ drives the system from the NFL state to
the LFL state, the flat portion of $\varepsilon({\bf p})$ determined
by Eq.~\eqref{tem} being destroyed at $T<T_{M}$ \cite{shagrep}. Thereupon
the factor $\eta$ vanishes, nullifying $\rho_0^c$ and strongly
reducing $\rho_0$. Since both $W(B)$ and $T^*(B)$ widths are
proportional $T$, imposition of the magnetic field causes a
step-like drop in the residual resistivity $\rho_0$. Consequently
two values of the residual resistivity must be introduced, namely
$\rho_0^{NFL}$ corresponding to the NFL state and $\rho_0^{LFL}$
corresponding to the LFL state induced upon application of the magnetic
field $B$.  It follows from these considerations that
$\rho_0^{NFL}>\rho_0^{LFL}$.  This conclusion agrees with the
experimental findings \cite{steg1,steg2,steg3}.
\begin{figure}[!ht]
\begin{center}
\includegraphics [width=0.47\textwidth]{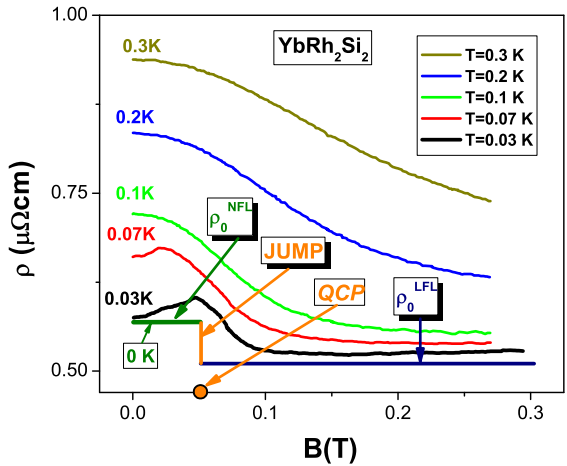}
\end{center}
\caption{(color online). Experimental results \cite{steg1} for the
longitudinal magnetoresistivity $\rho(T,B)$ of $\rm YbRh_2Si_2$
versus $B$ at various temperatures $T$. The maxima of the curves
for $T=0.03$ and $0.07$ K correspond to boundary points of the AF
ordered state shown in Fig.~\ref{fig2} {\bf b}. The solid lines
marked with $0$ K represent the schematic behavior of the residual
resistivity $\rho_0$ as a function of $B$. The arrows pointing to
the horizontal solid lines identify the residual resistivities
$\rho_0^{NFL}$ and $\rho_0^{LFL}$ in $\rm YbRh_2Si_2$. The jump of
$\rho_0$ occurs at the QCP identified by an arrow.} \label{fig3}
\end{figure}

Fig.~\ref{fig2} shows the $T-B$ phase diagram of $\rm YbRh_2Si_2$,
which maps faithfully onto the schematic phase diagram depicted in
Fig.~\ref{fig1}, except for the appearance of an AF phase at low
temperatures. As seen from Fig.~\ref{fig2}, at $T>T_N$ and $B=0$
the system in its NFL state, while the LFL phase prevails at low
temperature for magnetic fields beyond the critical value $B_{c0}$.
The respective residual resistivities are measured at
$\rho_0^{NFL}\simeq 0.55\,{\rm \mu\Omega cm}$ (NFL) and $\simeq
0.5\,{\rm \mu\Omega cm}$ (LFL) \cite{steg1}. As $T$ is lowered
through $T_N$ at $B=0$ the system enters the AF state via a
second-order phase transition.  Accordingly, we expect that the
residual resistivity does not change, remaining the same as that of
the NFL state, $\rho_0^{NFL}$.  On the other hand, under imposition
of an increasing $B$-field, the system moves from the NFL state to
the LFL state with the above value of $\rho_0^{LFL}$  At this point
it should be acknowledged that application of a weak magnetic field
is known to produce a positive classical contribution $\propto B^2$
to $\rho_0$ arising from orbital motion of carriers induced by the
Lorentz force.  When considering spin-orbit coupling in disordered
electron systems where electron motion is diffusive, the
magnetoresistivity may have both positive (weak localization) and
negative (weak anti-localization) signs \cite{larkin}. However, as
studied experimentally, $\rm YbRh_2Si_2$ is one of the purest
heavy-fermion metals.  Hence the applicable regime of electron
motion is ballistic rather than diffusive, both weak and anti-weak
localization scenarios are irrelevant, and one expects the
$B$-dependent correction to $\rho_0$ to be positive.  We therefore
conclude that the positive difference
$\rho_0^{c}=\rho_0^{NFL}-\rho_0^{LFL}$ comes from the contribution
related to the flat band. As seen from Fig.~\ref{fig3}, when the
system transits from the NFL state to the LFL state at fixed $T$
and under application of elevated magnetic fields $B$, the
step-like drop in its resistivity $\rho(T,B)$ becomes more
pronounced (see the experimental curves for $T=0.3, 0.2, 0.1$ K).
This behavior is a simple consequence of the fact that the width of
the crossover regime is proportional to $T$.  On zooming into the
vicinity of QCP shown in Fig.~\ref{fig3} (corresponding for example
to the experimental curves for $T=0.07$ and $0.03$ K), it may be
seen that the crossover width remains proportional to temperature,
ultimately shrinking to zero and leading to the abrupt jump in the
residual resistivity at $T=0$ when the system crosses the QCP at
$B=B_{c0}$.  In the same way, application of a magnetic field $B$
to $\rm CeCoIn_5$ causes a step-like drop in its residual
resistivity, as is in fact found experimentally \cite{pag1}. Based
on this reasoning, we expect that the higher the quality of both
$\rm CeCoIn_5$ and $\rm YbRh_2Si_2$ single crystals, the greater is
the ratio $\rho_0^{NFL}/\rho_0^{LFL}$, since the contribution
coming from the impurities diminishes and $\rho_0^{NFL}$ approaches
$\rho_0^{c}$. It is also expected from Eq.~\eqref{rho} that the
observed difference $\rho_0^c$ in the residual resistivities will
not show a marked dependence on the imperfection of the single
crystal unless the impurities destroy the flat band. Finally, we
point out that the jump of the magnetoresistivity at zero
temperature contradicts the usual behavior of Kondo systems, with
the width of the transition remaining finite at $T\to0$
\cite{steg2,s_k}. Moreover, the Kondo systems has nothing to do
with the dissymmetrical tunnelling conductance as a function of the
applied voltage $V$ that was predicted to emerge in such HF metals
with the flat band as $\rm CeCoIn_5$ and $\rm YbRh_2Si_2$
\cite{shagrep,shagd,shagpl}. Indeed, experimental observations have
revealed that the conductance is the dissymmetrical function of $V$
in both $\rm CeCoIn_5$ \cite{park} and $\rm YbRh_2Si_2$
\cite{stegc}.

The emergence of a flat band entails a change of the Hall
coefficient $R_H=\sigma_{xyz}/\sigma^2_{xx}$ \cite{norman,khodsp}.
In homogeneous matter at $B\to 0$ one has $\sigma_{xx}=\sigma/3$,
while $\sigma_{xyz}$ is recast to
\begin{equation}
\sigma_{xyz}={e^3\over 3\gamma^2}\int \left[{dz\over dp}\right]^2
{\partial n(z)\over
\partial z} dz,\label{sxyz}
\end{equation}
where $n(z)$ is the quasiparticle distribution function.  Far from
the QCP, these formulas lead to the standard result $R_H=1/Ne$,
whereas in the vicinity of the QCP, one finds $R_H=K/Ne$ with
$K\simeq 1.5$ \cite{khodsp}. We see then that the effective volume
of the Fermi sphere shrinks considerably at the QCP.  Importantly,
in the LFL state where the effective mass stays finite, the value
$K=1$ holds even quite close to the QCP.  As we have learned, the
width $W(B)$ tends to zero at the QCP, implying that the critical
behavior of $K$ at $T\to 0$ emerges abruptly, producing a jump in
the Hall coefficient, while the height of the jump remains finite.
It is instructive to consider the physics of this jump of $R_H$ in
the case of $\rm YbRh_2Si_2$.  At $T=0$, the critical magnetic
field $B_{c0}$ destroying the AF phase is determined by the
condition that the ground-state energy of the AF phase be equal to
the ground-state energy of the HF liquid in the LFL paramagnetic
state. Hence, at $B\to B_{c0}$ the N\'eel temperature $T_{N}$ tends
to zero. In the measurements of the Hall coefficient $R_H$ as a
function of $B$ performed in $\rm YbRh_2Si_2$
\cite{steg2,steg_cm,pash}, a jump is detected in $R_H$ as $T\to 0$
when the applied magnetic field reaches its critical value
$B=B_{c0}$ and then becomes infinitesimally higher at
$B=B_{c0}+\delta B$.  At $T=0$, application of the critical
magnetic field $B_{c0}$, which suppresses the AF phase whose Fermi
momentum is $p_F$, restores the LFL phase with a Fermi momentum
$p_f>p_F$. This occurs because the quasiparticle distribution
function becomes multiply connected and the number of mobile
electrons does not change \cite{shagrep}. The AF state can then be
viewed as having a ``small'' Fermi surface characterized by the
Fermi momentum $p_F$, whereas the LFL paramagnetic ground state at
$B>B_{c0}$ has a ``large'' Fermi surface with $p_f>p_F$. As a
result, the Hall coefficient experiences a sharp jump because
$R_H(B)\propto1/p_F^3$ in the AF phase and $R_H(B)\propto1/p_f^3$
in the paramagnetic phase. Assuming that $R_H(B)$ is a measure of
the Fermi momentum \cite{norman,pash} (as is the case with a simply
connected Fermi volume), we obtain \cite{shagrep,spa}
\begin{equation}
\frac {R_H(B=B_{c0}-\delta)}{R_H(B=B_{c0}+\delta)}\simeq1+3\frac
{p_f-p_F}{p_F}.\label{SL7}
\end{equation}
These observations are in excellent agreement with the experimental
facts collected on $\rm YbRh_2Si_2$ \cite{steg2,steg_cm}.

Violation of the Wiedemann-Franz law at the QCP in HF metals was
predicted and estimated a few years ago \cite{shagrep,VWF} and
recently observed \cite{steg3}.  Predictions of LFL theory fail in
the vicinity of a QCP where the effective mass $M^*$ diverges,
since the sp spectrum possesses a flat band at that point.  In a
once-standard scenario for such a QCP \cite{col1,col2}, the
divergence of the effective mass is attributed to vanishing of the
quasiparticle weight $Z$.  However, as already indicated, this
scenario is flawed \cite{khodz}.  We therefore employ a different
scenario for the QCP, in which the departure of the Lorenz number
$L$ from the Wiedemann-Franz value is associated with a
rearrangement of sp degrees of freedom leading to a flat band.
Within the quasiparticle paradigm, the relation between the Seebeck
thermodynamic coefficient $S$ and the conductivities $\sigma$ and
$\kappa$ has the form \cite{kin,ashcroft}
\begin{equation}
{\kappa(T)\over \sigma(T) T}+S^2(T)= {1\over e^2}{I_2(T)\over
I_0(T)}. \label{wfL}
\end{equation}
Here \begin{equation} S(T)={1\over e}{I_1(T)\over I_0(T)}  \  ,
\label{see}\end{equation}
with
\begin{equation} I_k(T)= -\int\!\!
\left({\epsilon(p)\over T}\right)^k \!\! \left({d\epsilon(p)\over
dp}\right)^2 \!\! \tau(\epsilon,T){\partial n(p)\over
\partial\epsilon(p)} d\upsilon , \label{ik}
\end{equation}
where $\tau$ is the collision time, $d\upsilon$ is the volume
element of momentum space, and $n(p)$ is given by Eq.~\eqref{HC2}
with $\epsilon=\varepsilon-\mu$. Overwhelming contributions to the
integrals $I_k$ come from a narrow vicinity $|\epsilon|\sim T$ of
the Fermi surface. In case of LFL, the Seebeck coefficient $S(T)$
vanishes linearly with $T$ at $T\to 0$. Then, the group velocity
can be factored out from the integrals \eqref{ik}. The same is true
for the collision time $\tau$, which at $T\to 0$ depends merely on
impurity scattering, and one obtains $I_1(T=0)=0$ and $I_2(T\to
0)/I_0(T\to 0)=\pi^2/3$. Inserting these results into
Eq.~\eqref{wfL}, we do find that the Wiedemann-Franz law holds,
even if several bands cross the Fermi surface simultaneously
\cite{kin}. On the other hand, taking into account the fact that
the reduction of the ratio $L/L_0$ occurs in the NFL state at the
QCP \cite{VWF}, we conclude that that the violation of the
Wiedemann-Franz law takes place in the narrow segment of the $T-B$
phase diagram displayed in Fig.~\ref{fig2} having width $W\to0$ at
$T\to0$.  In other words, at $T\to0$ the ratio $L/L_0$ becomes
abruptly $L/L_0\sim 0.9$ at $B/B_{c0}=1$, while $L/L_0=1$ at
$B/B_{c0}\neq1$ when the system is in its AF or LFL state shown in
Fig.~\ref{fig2}. This observation is in a good agreement with
experimental facts collected on $\rm YbRh_2Si_2$ \cite{steg3}. We
conclude that at $T\to0$, the WF law holds in the LFL state at
which the Fermi distribution function given by Eq. \eqref{HC2} is
reduced to the step function. The violation at $B=B_{c0}$ and at
$T\to0$ seen in $\rm YbRh_2Si_2$ thus suggests that a sharp Fermi
surface does exist at $B/B_{c0}\neq1$  but does not exist only at
$B/B_{c0}=1$ where the flat band emerges.

\begin{figure}[!ht]
\begin{center}
\includegraphics [width=0.47\textwidth]{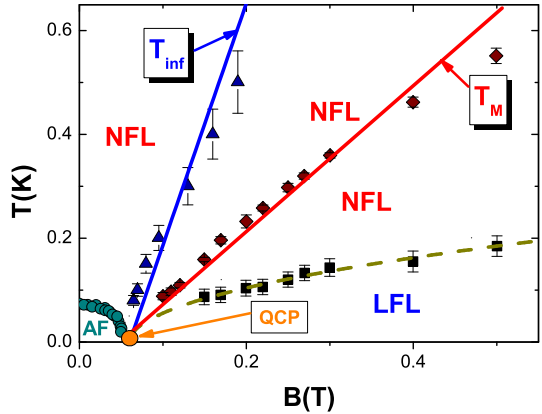}
\end{center}
\caption{(color online). Temperature versus magnetic field $T-B$
phase diagram for $\rm YbRh_2Si_2$. Solid circles represent the
boundary between AF and NFL states and the QCP shown by the arrow.
Solid squares refer to the boundary between NFL and LFL regimes
\cite{steg1,steg2} represented by the dashed line, which is
approximated by $(B-B_{c0})^{1/2}$ \cite{shagrep}. Diamonds mark
the maxima $T_M$ of the specific heat $C/T$ \cite{oes}, which are
approximated by $T_M\propto b_1(B-B_{c0})$, with $b_1$ a fitting
parameter \cite{shagrep}. Triangles close to the solid line refer
to the inflection points $T_{\rm inf}$ in the longitudinal
magnetoresistivity \cite{steg1,steg2}, while the solid line tracks
the function $T_{\rm inf}\propto b_2(B-B_{c0})$, with $b_2$ a
fitting parameter.} \label{fig4}
\end{figure}
Among other features, Fig.~\ref{fig4} includes results (solid
lines) for the characteristic temperatures $T_{\rm inf}(B)$ and
$T_M(B)$, which represent the positions of the kinks separating the
energy scales identified experimentally in
Refs.~\cite{steg1,steg2,oes}. The boundary between the NFL and LFL
phases is indicated by a dashed line, while AF labels the
antiferromagnetic phase; again the corresponding data are taken
from Refs.~\cite{steg1,steg2,oes}. It is seen that our calculations
are in accord with the experimental facts. In particular, we
conclude that the energy scales and the widths $W$ and $T^*$ are
reproduced by Eqs.~\eqref{fin1} and \eqref{TMB} and related to the
special points $T_{\rm inf}$ and $T_M$ associated with the
normalized effective mass $M^*_N$, which are marked with arrows in
the inset and main panel of Fig.~\ref{fig1} \cite{shagrep,scale}.

In summary, we have shown that imposition of a magnetic field on
$\rm YbRh_2Si_2$ leads to the emergence of the quantum critical
point at which a strong suppression of the residual resistivity
$\rho_0$ is accompanied both by a jump of the Hall resistivity and
a violation of the Wiedemann-Franz law. The close similarity
between the behaviors of the Hall coefficient $R_H$,
magnetoresistivity $\rho$, and Lorenz number $L$ at the QCP
indicates that all transport measures reflect the same underlying
physics, which unambiguously entails an interpretation of the QCP
as arising from a fermion condensation quantum phase transition
leading to the formation of a flat band.

We thank A. Alexandrov for fruitful discussions. This work was
supported by the U.S. DOE, Division of Chemical Sciences, Office of
Basic Energy Sciences and the Office of Energy Research, AFOSR, as
well as the McDonnell Center for the Space Sciences.

\end{document}